\documentclass[aps,prd,twocolumn,superscriptaddress,nofootinbib,preprintnumbers,floatfix,notitlepage]{revtex4-1}

\usepackage{graphicx}
\usepackage{amsmath,amsfonts,amssymb}
\usepackage{hyperref}
\usepackage{color}
\usepackage{verbatim}

\newcommand{\be}{\begin{equation}}
\newcommand{\ee}{\end{equation}}
\newcommand{\td}{{\rm d}}
\newcommand{\khat}{\hat{\boldsymbol{k}}}
\newcommand{\xhat}{\hat{\boldsymbol{x}}}

\begin{document}

\preprint{KCL-PH-TH/2020-36}

\title{Gravitational wave spectra from strongly supercooled phase transitions}

\author{Marek Lewicki}
\email{marek.lewicki@kcl.ac.uk}
\affiliation{Physics Department, King's College London, London WC2R 2LS, UK}
\affiliation{Faculty of Physics, University of Warsaw ul.\ Pasteura 5, 02-093 Warsaw, Poland}
\author{Ville Vaskonen}
\email{ville.vaskonen@kcl.ac.uk}
\affiliation{Physics Department, King's College London, London WC2R 2LS, UK}
\affiliation{NICPB, R\"avala 10, 10143 Tallinn, Estonia}

\begin{abstract}
We study gravitational wave (GW) production in strongly supercooled cosmological phase transitions, taking particular care of models featuring a complex scalar field with a U$(1)$ symmetric potential. We perform lattice simulations of two-bubble collisions to properly model the scalar field gradients, and compute the GW spectrum sourced by them using the thin-wall approximation in many-bubble simulations. We find that in the U$(1)$ symmetric case the low-frequency spectrum is $\propto\omega$ whereas for a real scalar field it is $\propto\omega^3$. In both cases the spectrum decays as $\omega^{-2}$ at high frequencies.
\end{abstract}

\maketitle

\section{Introduction}

The direct detection of gravitational waves (GWs) from a binary black hole merger by LIGO~\cite{Abbott:2016blz} marked the dawn of a new era in astrophysics and cosmology. In the next decades various experiments will probe GWs in a wide range of frequencies~\cite{Janssen:2014dka,Audley:2017drz,Graham:2016plp,Graham:2017pmn,Badurina:2019hst,Bertoldi:2019tck,Punturo:2010zz,Hild:2010id}. In addition to the astrophysical GW sources, such as compact object binaries, these experiments will probe also cosmological GW backgrounds providing a unique probe of the early Universe, as, unlike electromagnetic signals, GWs can propagate freely from the very beginning of the Universe.

Cosmological first-order phase transitions constitute one possible source of GWs from the early Universe~\cite{Witten:1984rs}, which can be probed by the upcoming GW experiments~\cite{Caprini:2015zlo,Caprini:2019egz}. In a first-order phase transition the false vacuum is separated from the true vacuum by a barrier as the transition proceeds. As a result the unstable vacuum decays through nucleation of bubbles, corresponding to the field trapped in the false vacuum tunnelling through the barrier~\cite{Coleman:1977py,Callan:1977pt,Linde:1981zj}. After nucleation these bubbles grow until they collide, eventually converting the whole Hubble volume into the new phase. 

The bubbles typically grow by many orders of magnitude between their nucleation and collisions, releasing a lot of energy. This energy goes into gradient and kinetic energy of the bubble walls and into motion of the plasma as the particles in the plasma interact with the bubble wall. GWs from a phase transition are sourced by the scalar field gradients~\cite{Kosowsky:1992vn} and motions in the plasma~\cite{Kamionkowski:1993fg}. For very strongly supercooled phase transitions the plasma friction can be negligible. In this case the bubble walls reach velocities near the speed of light before they collide~\cite{Bodeker:2009qy,Bodeker:2017cim} and the GW signal is dominated by the scalar field gradients~\cite{Ellis:2019oqb}.

The GW signal from scalar field gradients was first calculated in the envelope approximation in Ref.~\cite{Kosowsky:1992vn}. In this approximation the bubble walls are treated as thin shells that disappear in the collisions. The resulting GW spectrum is a broken power-law that at low frequencies grows as $\omega^3$ and at high frequencies decays as $\omega^{-1}$~\cite{Huber:2008hg,Weir:2016tov,Konstandin:2017sat}. In Refs.~\cite{Jinno:2017fby,Konstandin:2017sat} the envelope approximation was extended in order to model colliding fluid shells. In this so called bulk flow model the bubble wall energy was assumed to decay as $R^{-2}$ after the collision as a function of the bubble radius $R$. The bulk flow approximation results in a GW spectrum that turns from $\omega^{1}$ behaviour to $\omega^{-2}$ at around the same frequency at which the spectrum in the envelope approximation peaks. 

Recently the GW spectrum was calculated in 3D lattice simulations~\cite{Child:2012qg,Cutting:2018tjt,Cutting:2020nla}. 
These simulations are very difficult because of the large separation between the characteristic length scales in the problem, that is the size of the growing bubble and its thinning wall. Due to numerical limitations in such simulations it is impossible account for realistically large bubble wall velocities. However, the GW spectrum can still be computed. At high frequencies it was found that the spectrum lies somewhere between the envelope and bulk flow approximations. The low-frequency behaviour of the spectrum is especially difficult to resolve and a $\omega^3$ behaviour is typically assumed.

In this paper we approximate the GW spectrum from strongly supercooled phase transitions by first studying the scaling of the gradient energy in two-bubble collisions by lattice simulations and then calculating the GW spectrum by performing many-bubble simulations in thin-wall approximation. In this way we can efficiently perform large simulations with realistic behaviour of the GW source. In contrast to the earlier works, we study also the case of a complex scalar field with a U$(1)$ symmetric potential which is often realized in particle physics models~\footnote{See e.g. Refs.~\cite{Huber:2015znp,Jinno:2016knw,Iso:2017uuu,Demidov:2017lzf,Hashino:2018zsi,Marzo:2018nov,Miura:2018dsy,Azatov:2019png} for studies where GW signal was studied in these kind of models.}. We find that in this case the gradient energy quickly reaches an $R^{-2}$ scaling after the collision, whereas for a real scalar field we find that the decay is much faster. In the former case we find that the GW spectrum is near the bulk flow result, and in the latter case we find a GW spectrum that grows as $\omega^3$ at low frequencies and decays as $\omega^{-2}$ at high frequencies.

\section{Bubble collisions}

We begin by studying collisions of two complex scalar field bubbles. In order for the scalar field gradients to be the dominant source of GWs the phase transition has to be severely supercooled~\cite{Ellis:2019oqb}. Typically this can not be realized in models based on polynomial potentials~\cite{Ellis:2018mja}, but in models that are classically scale invariant a prolonged period of supercooling is possible~\cite{Randall:2006py,Konstandin:2011dr,Konstandin:2011ds,Jinno:2016knw,Iso:2017uuu,vonHarling:2017yew,Kobakhidze:2017mru,Marzola:2017jzl,Prokopec:2018tnq,Hambye:2018qjv,Marzo:2018nov,Baratella:2018pxi,Bruggisser:2018mrt,Aoki:2019mlt,DelleRose:2019pgi,Fujikura:2019oyi,Wang:2020jrd}. In these models the symmetry breaking originates from radiative corrections~\cite{Coleman:1973jx} and finite temperature effects give raise to a potential energy barrier between the symmetric and the symmetry-breaking minima. The one-loop effective potential is of the form
\be \label{eq:pot0}
V(\phi) = B \phi^4 \left[\ln\left(\frac{|\phi|^2}{v^2}\right) - \frac{1}{4} \right] + C T^2 |\phi|^2
\ee
where $B$ and $C$ are dimensionless constants that depend on the couplings of the scalar field $\phi$~(see e.g.~\cite{Marzola:2017jzl}), $v$ is the vacuum expectation value of $|\phi|$ and $T$ denotes the temperature of the plasma. Motivated by this, we consider logarithmic potential
\be \label{eq:pot}
\frac{V(\phi)}{\Delta V} = 1 \!+\! \kappa \frac{|\phi|^2}{v^2} \!+\! \frac{|\phi|^4}{v^4} \left[(\kappa\!+\!2)\log\!\left(\frac{|\phi|^2}{v^2}\right) \!-\! (\kappa\!+\!1)\right] ,
\ee
where $\kappa$ is a dimensionless parameter. This potential is U$(1)$ symmetric and its global minimum lies at $|\phi| = v$ where $V(|\phi| = v) = 0$. For $\kappa>0$ the point $\phi=0$ is a local minimum with $V(0) = \Delta V$. The parameters $B$ and $C$ of Eq.~\eqref{eq:pot0} are related to $\kappa$ and $\Delta V$ via
\be
B = \frac{(\kappa+2)\Delta V}{v^4} \,,\quad C = \frac{\kappa\Delta V}{T^2 v^2} \,.
\ee

The radial initial profile of the modulus $|\phi|$ for an $O(4)$ symmetric bubble is obtained as the solution of 
\be \label{eq:O4initialbubble}
\partial_r^2 |\phi| + \frac{3}{r} \partial_r |\phi| = \frac{\td V}{\td|\phi|}
\ee
with boundary conditions $\partial_r\phi = 0$ at $r=0$ and $\phi\to 0$ at $r\to \infty$. Due to the $U(1)$ symmetry of the potential every bubble will be nucleated with a complex phase $\varphi$ of the field $\phi$ chosen from the range $\varphi\in [0,2\pi[$ with equal probability.

We assume that the phase transition finishes within a Hubble time and therefore neglect the background expansion. Collision of two initially $O(4)$ symmetric scalar field bubbles is $O(1,2)$ symmetric, and it is convenient to define new coordinates~\cite{Hawking:1982ga} $(s,z,\psi,\theta)$ by $\tan\theta = x/y$
\be \label{eq:spsi}
\begin{aligned}
&t = s\cosh\psi \,, \quad r = s\sinh\psi \,, \qquad {\rm for}\,\, t\geq r \,, \\
&t = s\sinh\psi \,, \quad r = s\cosh\psi \,, \qquad {\rm for}\,\, t<r \,,
\end{aligned}
\ee
where $r^2=x^2+y^2$. The bubbles lie at the $z$ axis. The Klein-Gordon equation for the real ($X=R$) and imaginary ($X=I$) parts of the field in these coordinates simplify to
\be
\pm\partial_s^2 \phi_X \pm \frac{2}{s} \partial_s \phi_X - \partial_z^2 \phi_X = -\frac{\td V}{\td\phi_X} \,,
\ee
where $+$ and $-$ signs correspond to the regions $t\geq r$ and $t<r$, respectively. The collision of bubbles occurs in the region $t\geq r$ where we solve the above equation numerically. In the region $t<r$ the evolution is given by analytical continuation of the initial bubble solution
\be
\phi(s,z) = \sum_j \phi_0\left[\sqrt{s^2+(z-z_j)^2}\right] \,,
\ee
where $z_j$ denotes the position of the bubble $j$.

\begin{figure*}
\centering
\includegraphics[width=\textwidth]{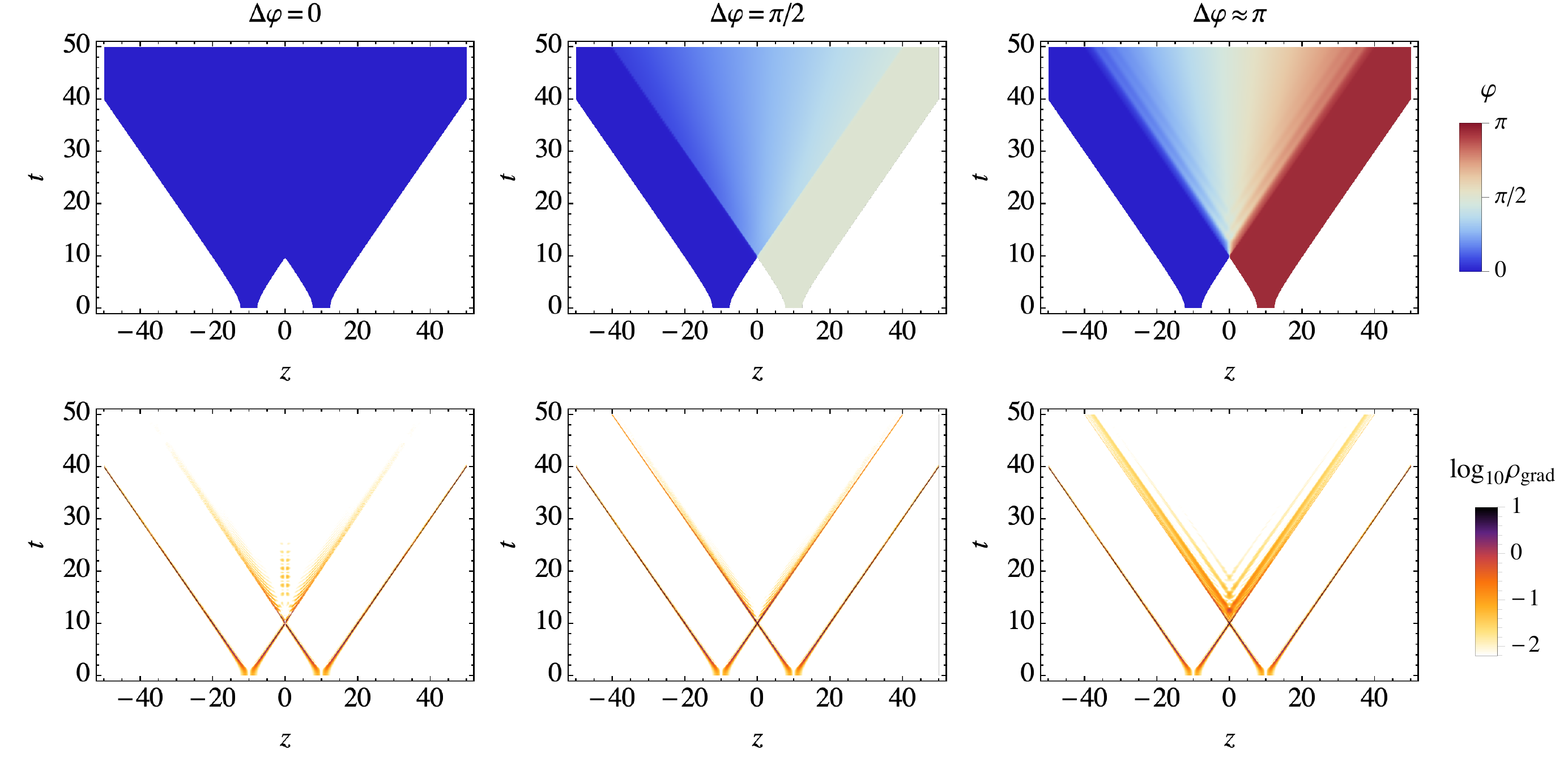}
\vspace{-6mm}
\caption{Collision of two complex scalar field bubbles in the simulation units. Each column corresponds to one simulation. The color coding in the upper panels shows the complex phase of the scalar field, whereas in the lower panels it shows the gradient energy density. White color in the upper panel indicates the region where $\phi \approx 0$.}
\label{fig:collisions}
\end{figure*}

In Fig.~\ref{fig:collisions} we show the result from two-bubble collision in three values of the phase difference between the colliding bubbles: $\Delta \varphi=0$, $\Delta \varphi=\pi/2$ and $\Delta \varphi\approx \pi$~\footnote{Taking exactly $\Delta\varphi = \pi$ we would form a stable domain wall in the collision due to $Z_2$ symmetry of that very particular configuration.}. The numerical lattice calculation is performed in dimensionless variables, obtained by scaling $\phi \to \phi/v$ and $x^{\mu} \to \sqrt{\Delta V} x^{\mu}/v$, and the results in Fig.~\ref{fig:collisions} are shown in the simulation units. We see that the bubble walls quickly accelerate after nucleation, approaching velocities near the speed of light before their collision.\footnote{In Fig.~\ref{fig:collisions} curves of the form $t=\pm z + C$, where C is a constant, are lightlike.} After the collision we see that a sharp phase wall continues to propagate with a constant velocity near the speed of light. This can also be seen in the lower panels, where we show the evolution of the gradient energy density of the scalar field, $\rho_{\rm grad} = |\partial_z \phi|^2/2$. It is also clear from these plots that the energy loss of the gradients after the collision is much faster in the case where the bubbles have equal complex phases effectively corresponding to the case with a real scalar field.

\begin{figure}
\centering
\includegraphics[width=0.98\columnwidth]{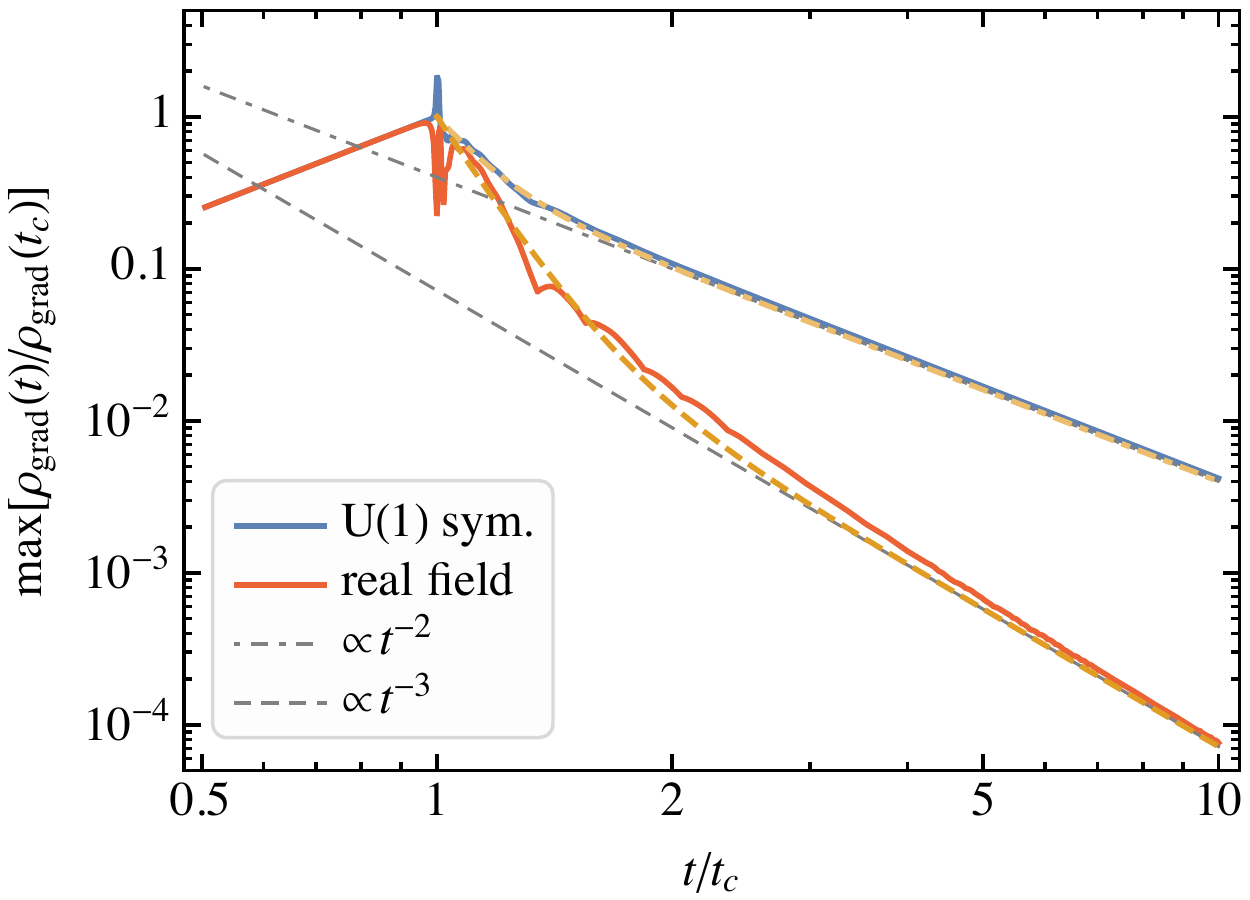}
\vspace{-2mm}
\caption{Evolution of the maximal gradient energy density in a bubble collision. The blue solid line shows the average from simulations with different values of the complex phase difference between the colliding bubbles, and the red solid line shows the case where the bubbles have equal complex phase. The yellow dotdashed and dashed lines show the broken power-law approximations for the evolutions after the collision.}
\label{fig:decay}
\end{figure} 

In Fig.~\ref{fig:decay} we show by the blue solid line the time evolution of the maximum of gradient energy density averaged over various values of the phase difference $\Delta\varphi$. We obtain the case of a real scalar field from our analysis by taking only the case $\Delta\varphi = 0$. This is shown by the red solid line. The collision happens at $t=t_c$.  Before the collision the total released energy scales as $E_{\rm rel}\propto t^3$, the surface area as $A\propto t^2$ and the wall thickness as $L\propto 1/t$ due to increase in the Lorentz factor of the wall. Therefore the energy density at the wall scales as $\rho_{\rm grad} \propto E_{\rm rel}/(AL) \propto t^2$, which is what we see also in Fig.~\ref{fig:decay}. After the collision, if the bubble wall velocity is constant and the energy remains localized at the bubble wall, its gradient energy density scales as $\rho_{\rm grad} \propto t^{-2}$ as $E_{\rm rel}$ and $L$ remain constant. From Fig.~\ref{fig:decay} we see that this is not a perfect description. In reality some of the energy is spread into the collided volume and therefore the maximal gradient energy density decays faster. In the case of a real scalar field we see that the scaling of gradient energy density reaches $\propto t^{-3}$ behaviour after the collision, while in the U$(1)$ symmetric case it scales significantly slower, $\propto t^{-2}$. This is a consequence of the phase difference between the colliding bubbles that after the collision for a long time continues to propagate as a sharp phase domain wall.

The results shown in Figs.~\ref{fig:collisions} and \ref{fig:decay} are from simulations with $\kappa=0.2$, but we have checked that the scaling of the gradient energy density after the collision is not sensitive to the value of $\kappa$. We have also checked that the same scaling results hold in the case of a polynomial potential~\footnote{In the class of potentials we focus on false vacuum trapping in the collision typically does not occur. If it did that would change the result, as most of the gradient energy would remain around the collision point~\cite{Jinno:2019bxw,Lewicki:2019gmv}.}. In addition, we have run simulations with different values for the initial bubble separation $d$, that in Figs.~\ref{fig:collisions} and \ref{fig:decay} was set to $d=20$ (in the simulation units). Our results from simulations with $d\in [8,200]$ indicate that the bubble separation dictating that the $\gamma$ factor of the wall at collision and its energy does not change the scaling behaviour.

\section{Production of gravitational waves}
\label{sec:gws}

Next we will study the production of GWs from scalar field gradients using the scaling results of the bubble wall energy obtained in the previous section. Following Ref.~\cite{Weinberg:1972kfs}, the total energy spectrum in a direction $\khat$ at frequency $\omega$ of the GWs emitted in the phase transition is given by
\be \label{eq:dEkdw}
\frac{\td E}{\td\Omega_k\td \omega} = 2G\omega^2 \Lambda_{ijlm}(\khat) T_{ij}^*(\khat,\omega) T_{lm}(\khat,\omega) \,,
\ee
where $\Lambda_{ijlm}$ is the projection tensor,
\be
\begin{aligned}
\Lambda_{ijlm}(\khat) =& \,\delta_{il}\delta_{jm} -2\delta_{il}\khat_j\khat_m + \frac{1}{2}\khat_i \khat_j \khat_l \khat_m \\&- \frac{1}{2} \delta_{ij}\delta_{lm} + \frac{1}{2}\delta_{ij}\khat_l\khat_m + \frac{1}{2}\delta_{lm}\khat_i\khat_j \,,
\end{aligned}
\ee
and $T_{ij}$ is the traceless part of the stress energy tensor,
\be \label{eq:Tij}
T_{ij}(\khat,\omega) = \frac{1}{2\pi} \int \td t\, \td^3 x \,e^{i\omega(t-\khat\cdot\boldsymbol{x})} \,\partial_i \phi \partial_j \phi^* \,.
\ee

In the thin-wall limit, the gradient energy carried by an uncollided element of the bubble wall at solid angle $\td \Omega_x$ can be approximated as~\cite{Kosowsky:1992vn}
\be \label{eq:gradE}
\begin{aligned}
\td\Omega_x &\int \td r\, r^2 e^{-i\omega \khat\cdot\boldsymbol{x}} \partial_i \phi \partial_j \phi^* \\
&\approx \td\Omega_x\, \xhat_i \xhat_j \, \frac{R_n^3 \Delta V}{3} e^{-i\omega \khat\cdot(\boldsymbol{x}_n+R_n\xhat)} \,,
\end{aligned}
\ee
where $\boldsymbol{x}_n$ denotes the position vector of the bubble center, $\xhat$ is a unit vector that points from the centre of the bubble in the direction $\td \Omega$ and $R_n \approx t-t_n$ is the radius of the bubble that nucleated at time $t_n$. After the element of the bubble wall at $\td\Omega_x$ has collided with another bubble, its energy starts to decrease. Assuming that the velocity of the wall element does not change in the collision, the scaling of the energy can be accounted by multiplying Eq.~\eqref{eq:gradE} by a function $f(t)$ which depends on the time $t_{n,c} = t_{n,c}(\xhat)$ when the wall element collides with another bubble. Before the collision $f(t<t_{n,c}) = 1$, and the envelope approximation corresponds to taking $f(t>t_{n,c}) = 0$. Assuming instead that the bubble wall loses energy $\propto R^{-2}$ we get the bulk flow approximation~\cite{Jinno:2017fby,Konstandin:2017sat} where $f(t>t_{n,c}) = [(t_{n,c}-t_n)/(t-t_n)]^3$.

On the basis of the results shown in Fig.~\ref{fig:decay}, we find that the decay of the maximum of the gradient energy density can be approximated as a broken power-law, that changes from an $\propto R^{-\xi_1}$ behaviour to $\propto R^{-\xi_2}$. In the thin wall approximation, the bubble wall energy is simply $\td E_{\rm wall} \propto \td \Omega_x R^2 L \max[\rho_{\rm grad}]$, where the bubble wall width $L$ after the collision is constant, and therefore
\be
f(t>t_{n,c}) = \sum_{j=1}^2 b_j \left(\frac{t_{n,c}-t_n}{t-t_n}\right)^{\xi_j+1} \,.
\ee
For the U$(1)$ symmetric case we use $b_1=0.6$, $b_2=0.4$, $\xi_1=8$, $\xi_2=2$, and in the case of a real scalar field we use $b_1=0.93$, $b_2=0.07$, $\xi_1=8$, $\xi_2=3$. These approximations are shown in Fig.~\ref{fig:decay} with the yellow dotdashed and dashed lines, respectively.

The contribution from $N$ bubbles on the traceless part of the stress energy tensor~\eqref{eq:Tij} can now be written as
\be
\begin{aligned}
T_{ij}(\khat,\omega) \approx &\frac{\Delta V}{6\pi} \sum_{n=1}^N \int_{t_n} \td t\, \td\Omega_x\, \xhat_i \xhat_j \\ &\times f(t,t_{n,c}) R_n^3 \,e^{i\omega [t - \khat\cdot(\boldsymbol{x}_n+R_n\xhat)]} \,.
\end{aligned}
\ee
We can rotate the coordinate system such that a given $\khat = \khat(\phi_k,\theta_k)$ after the rotation points to $z$ direction, $\khat \to \khat' = (0,0,1)$. Then, the projection in Eq.~\eqref{eq:dEkdw} simplifies to~\cite{Huber:2008hg}
\be
\begin{aligned}
\frac{\td E}{\td\Omega_k\td \omega} 
&= G\omega^3\left(|T_{xx} - T_{yy}|^2 + 2|T_{xy}|^2 + 2|T_{yx}|^2\right) \\ 
&= G\Delta V^2\omega^2 \left(|C_+|^2 + |C_\times|^2\right)\,,
\end{aligned}
\ee
where
\be \label{eq:Cpc}
\begin{aligned}
C_{+,\times}(\khat',\omega) \approx &\frac{1}{6\pi} \sum_{n=1}^N \int_{t_n} \td t\, \td \Omega_x\, \sin^2\theta_x'\, g_{+,\times}(\phi_x') \\ &\times f(t,t_{n,c}) R_n^3 \,e^{i\omega (t - z_n' - R_n\cos\theta_x')} \,,
\end{aligned}
\ee
with $g_+(\phi_x') = \cos(2\phi_x')$ and $g_\times(\phi_x') = \sin(2\phi_x')$. The spatial angles in the rotated coordinate system are
\be
\begin{aligned}
&\tan\phi_x' = \frac{\sin\theta_x\sin(\phi_x-\phi_k)}{\cos\theta_k \sin\theta_x \cos(\phi_x-\phi_k) - \sin\theta_k \cos\theta_x} \,, \\
&\cos\theta_x' = \sin\theta_k \sin\theta_x \cos(\phi_x-\phi_k) + \cos\theta_k \cos\theta_x\,.
\end{aligned}
\ee

We consider the exponential bubble nucleation rate per unit volume, $\Gamma \propto e^{\beta t}$. The parameter $\beta$, with mass dimension 1, sets the time and length scale of the transition. The abundance of GWs produced in bubble collisions in a logarithmic frequency interval is then
\be
\Omega_{\rm GW}(\omega) \equiv \frac{1}{E_{\rm tot}}\frac{\td E}{\td\ln\omega} = \left(\frac{H}{\beta}\right)^2\left(\frac{\alpha}{1+\alpha}\right)^2 S(\omega) \,,
\ee
where $\alpha = \Delta V/(\rho_{\rm tot} - \Delta V)$ characterizes the strength of the transition, $H^2 = 8\pi G\rho_{\rm tot}/3$ is the Hubble rate, and 
\be \label{eq:S}
S(\omega) = \left(\frac{\omega}{\beta}\right)^3 \frac{3\beta^5}{8\pi V_s} \int \td\Omega_k \left( |C_+|^2 + |C_\times|^2 \right),
\ee
gives the spectral shape of the GW background. The volume over which $\Omega_{\rm GW}$ is averaged is denoted by $V_s$. We note that $\int\!\td\Omega_k \left(|C_+|^2\!+\!|C_\times|^2\right) \propto V_s/\beta^5$. Next we calculate the $S(\omega)$ function numerically.

\section{Gravitational wave spectrum}

\begin{figure}
\centering
\includegraphics[width=0.98\columnwidth]{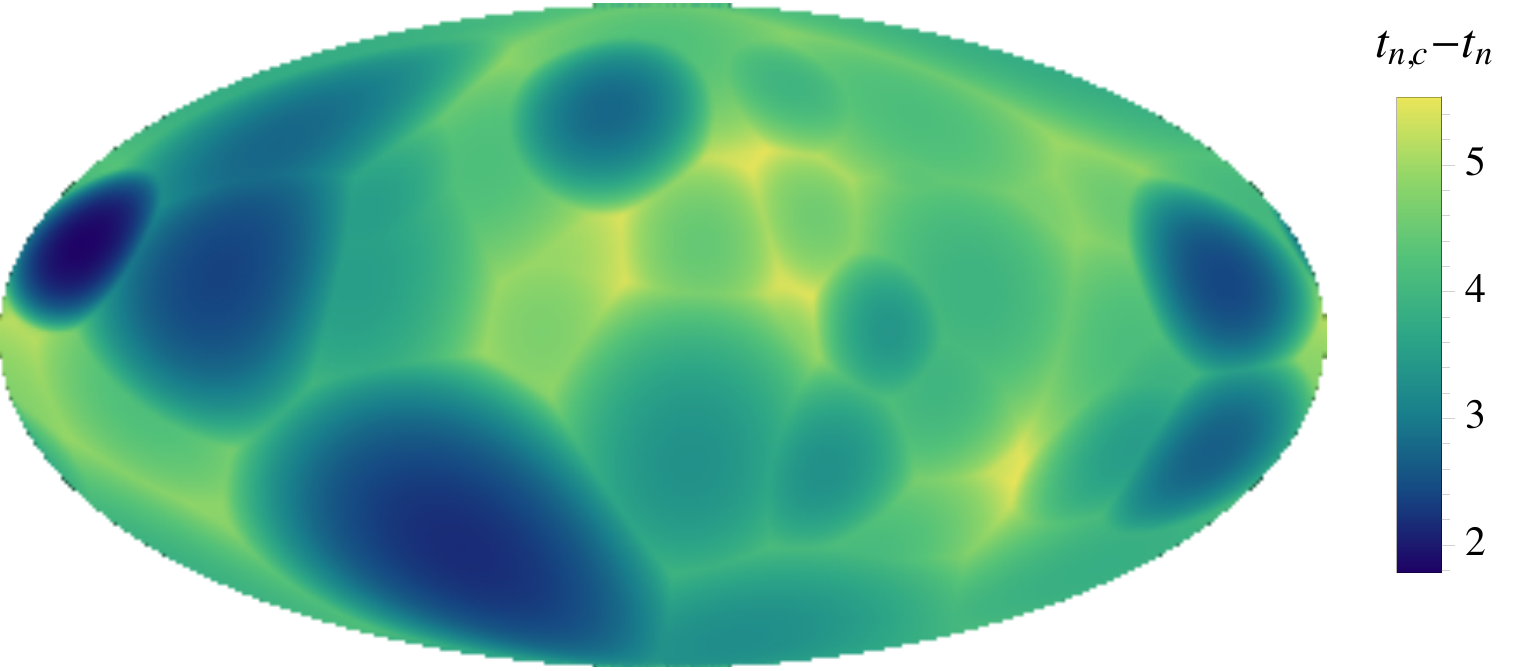}
\caption{Mollweide projection of a bubble surface in a many-bubble simulation. The color coding indicates the time when the given part of the bubble surface collided for the first time with another bubble.}
\label{fig:mollweide}
\end{figure}

In order to determine the spectral shape of the GW signal we simulate the phase transition by nucleating bubbles according to the rate $\Gamma \propto e^{\beta t}$ inside a cubic simulation volume with periodic boundary conditions. We neglect the initial bubble sizes, assume that their wall is infinitesimally thin and evolve the bubble radii as $R_n = t-t_n$. We generate points on the surface of a bubble and find the time $t_{n,c}$ when each of these points collides with another bubble surface by the bisection method~\cite{Burden1997}. As an example, in Fig.~\ref{fig:mollweide} we show the surface of the first bubble that nucleated with the color coding indicating the collision time $t_{n,c}$. Once the collision times are known, we can simply integrate the functions $C_{+,\times}$, and finally compute the GW spectrum~\eqref{eq:S}. 

The time integral in Eq.~\eqref{eq:Cpc} can be evaluated analytically if $f(t,t_{n,c})$ is a (broken) power-law. We perform the remaining integrals over $\khat$ and $\xhat$ directions numerically. As the simulation volume is not spherically symmetric we calculate the spectrum only for 6 $\khat$ directions that correspond to the normal vectors of the cube faces. In order to accurately determine the GW spectrum, we calculate the average spectrum over multiple simulations.\footnote{Averaging many smaller simulations is equivalent to running one much larger simulation but much more easily parallelised.}

\begin{figure}
\centering
\includegraphics[width=0.98\columnwidth]{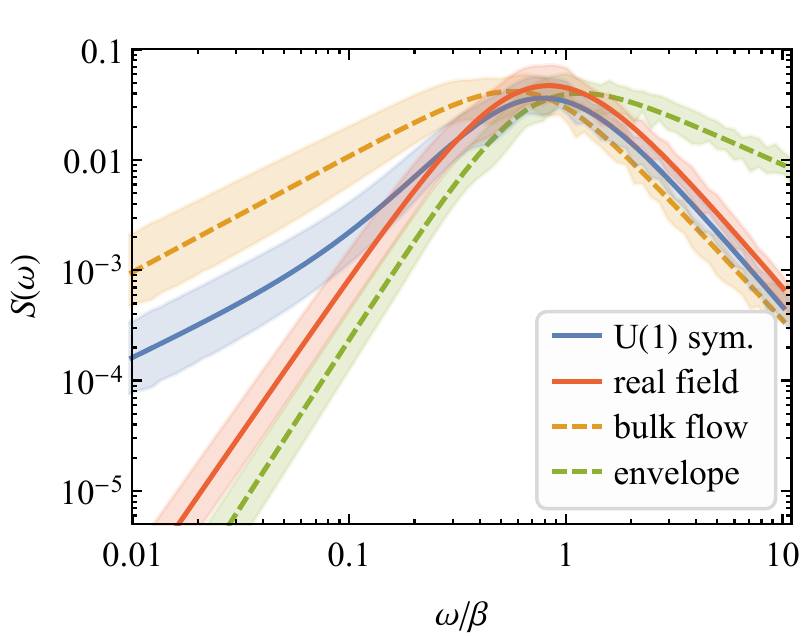}
\vspace{-1mm}
\caption{GW spectrum sourced by scalar field gradients. The solid curves show the spectral shape fit~\eqref{eq:fit} in the U$(1)$ symmetric case (blue) and in the case that all bubbles have equal complex phase (red) corresponding to a situation with just a real scalar field. The green and yellow dashed curves show the spectrum in the envelope and bulk flow approximations. The shaded bands indicate the variance from averaging over 40 simulations.}
\label{fig:gws}
\end{figure}

In Fig.~\ref{fig:gws} the red and blue solid curves show, respectively, our result for the spectral shape of the GW spectrum in the U$(1)$ symmetric case, and the case where all bubbles have equal complex phase, $\Delta\varphi=0$, which corresponds to having a real scalar field. The green and yellow dashed curves, for comparison, the result in the envelope and bulk flow approximations. The results are obtained by taking geometric mean over 40 simulations with around 200 bubbles each, while the error bands show the corresponding geometric standard deviations.

We calculate a broken power-law fit to the spectrum parametrized as
\be \label{eq:fit}
S_{\rm fit}(\omega) \!=\! \left[\frac{1\!+\!\left(\frac{\omega}{\omega_d}\right)^{d-a}}{1\!+\!\left(\frac{\bar\omega}{\omega_d}\right)^{d-a}}  \right] \!\frac{A\,(a+b)^c}{\left[b \left(\frac{\omega}{\bar\omega}\right)^{-a/c} \!+\! a \left(\frac{\omega}{\bar\omega}\right)^{b/c}\right]^c} \,,
\ee
where $A$ and $\bar\omega$ correspond to the peak amplitude and frequency, $c$ determines the width of the peak, and $a,b>0$ are the low- and high-frequency slopes near the peak of the spectrum. The first bracket parametrizes the change in the low-frequency slope which in the U$(1)$ symmetric case resembles the envelope result shortly after collision with emission from a slowly decaying gradient dominating later on. In the other cases we take $d=a$ for which the first bracket in~\eqref{eq:fit} gives 1.\footnote{Similar parametrization is used e.g. in Ref.~\cite{Cutting:2018tjt}. Instead, for example in Ref.~\cite{Konstandin:2017sat} the parameter $c$ is neglected.} The parameter values of the fit are shown in Table~\ref{table:fit}.

Finally, the present day GW spectrum can be obtained by simple red-shifting as~\cite{Kamionkowski:1993fg}
\be
\begin{aligned}
&\Omega_{{\rm GW},0} = \frac{1.67\!\times\!10^{-5}}{h^2} \left(\frac{H}{\beta}\right)^{\!2}\left(\frac{\alpha}{1+\alpha}\right)^{\!2} \!\left(\frac{100}{g_*}\right)^{\!\frac13}\! S(\omega) \,, \\
&\bar \omega_0 = h_* \left(\frac{\beta}{H}\right) \left(\frac{\bar\omega}{\beta}\right) \,,
\end{aligned}
\ee
where $h = 0.674$~\cite{Aghanim:2018eyx} denotes the dimensionless Hubble parameter and 
\be
h_* = 1.65\times 10^{-5}\,{\rm Hz}\, \left(\frac{T_*}{100\,{\rm GeV}}\right) \left(\frac{g_*}{100}\right)^{\frac16}
\ee 
is the inverse Hubble time at the transition redshifted to today\footnote{For the red-shifting we assume standard radiation dominated expansion up to the matter-radiation equality. For a review of possible deviations from this see Ref.~\cite{Allahverdi:2020bys}.}. The transition temperature is denoted by $T_*$ and the effective number of relativistic degrees of freedom at the temperature $T_*$ by $g_*$. At scales larger than the horizon scale at the time of the transition the spectrum scales as $\omega^3$ because the source is diluted by the Hubble expansion~\cite{Caprini:2009fx,Cai:2019cdl}. At the present time this corresponds to frequencies $\omega < h_*/(2\pi)$.

\begin{table}
\centering
\begin{tabular}{ p{1.4cm} p{0.8cm} p{0.8cm} p{0.8cm} p{0.8cm} p{0.8cm} p{0.8cm} p{0.8cm} }
\hline\hline
& $100 A$ & $\bar\omega/\beta$ & $\omega_d/\beta$ & $a$ & $b$ & $c$ & $d$\\
\hline
U$(1)$ sym. & 3.63 & 0.81 & 0.13 & 2.54 & 2.24 & 2.30 & 0.93 \\
real field & 4.71 & 0.83 & - & 2.84 & 2.29 & 2.52 & $a$ \\
bulk flow & 4.318 & 0.56 & - & 1.05 & 2.10 & 1.16 & $a$ \\
envelope & 4.02 & 1.16 & - & 3.03 & 0.88 & 1.55 & $a$ \\
\hline\hline
\end{tabular}
\caption{Fitted values of the parametrization~\eqref{eq:fit}.}
\label{table:fit}
\end{table}

\section{Conclusions}

We studied the GW spectrum produced in a strongly supercooled phase transition. We started from a lattice simulation of two-bubble collisions in order to model the evolution of the scalar field gradients which source GWs. We used a logarithmic potential typical for very strong phase transitions, and considered the impact of a complex scalar potential possessing an $U(1)$ symmetry. We then simulated the production of GWs assuming that after collisions the gradients will continue moving at velocities near the speed of light losing energy according to the lattice results.

We found that the collision fronts disappear much more slowly in collisions of bubbles with different complex phases than in the case where the phases are equal. Therefore in the $U(1)$ symmetric case the GW source is decaying significantly slower and the resulting GW spectrum is less steep at low frequencies than in the case of a real scalar field. Our results for the GW spectra are shown in Fig.~\ref{fig:gws}, from which we see that the low-frequency behaviours are in these cases very different: In the case of a real scalar field the low-frequency spectrum resembles the envelope approximation result $\propto \omega^{3}$, whereas in the U$(1)$ symmetric case it is closer to the bulk flow result $\propto \omega$. In both cases the high-frequency power-law is $\omega^{-2}$. We provided a simple broken power-law fits to these spectra, convenient for phenomenological studies.

Our treatment accounts only for the gradients at the bubble wall. Since the gradient energy is concentrated where the bubble wall would propagate also after the collisions, accounting for the field gradients in detail should result only in minor changes in the GW spectrum. However, to check this detailed lattice simulations are needed.

\begin{acknowledgments}
This work was supported by the UK STFC Grant ST/P000258/1. ML was partly supported by the Polish National Science Center grant 2018/31/D/ST2/02048 and VV by the Estonian Research Council grant PRG803.
\end{acknowledgments}

\bibliography{PBH}

\begin{thebibliography}{60}%
\makeatletter
\providecommand \@ifxundefined [1]{%
 \@ifx{#1\undefined}
}%
\providecommand \@ifnum [1]{%
 \ifnum #1\expandafter \@firstoftwo
 \else \expandafter \@secondoftwo
 \fi
}%
\providecommand \@ifx [1]{%
 \ifx #1\expandafter \@firstoftwo
 \else \expandafter \@secondoftwo
 \fi
}%
\providecommand \natexlab [1]{#1}%
\providecommand \enquote  [1]{``#1''}%
\providecommand \bibnamefont  [1]{#1}%
\providecommand \bibfnamefont [1]{#1}%
\providecommand \citenamefont [1]{#1}%
\providecommand \href@noop [0]{\@secondoftwo}%
\providecommand \href [0]{\begingroup \@sanitize@url \@href}%
\providecommand \@href[1]{\@@startlink{#1}\@@href}%
\providecommand \@@href[1]{\endgroup#1\@@endlink}%
\providecommand \@sanitize@url [0]{\catcode `\\12\catcode `\$12\catcode
  `\&12\catcode `\#12\catcode `\^12\catcode `\_12\catcode `\%12\relax}%
\providecommand \@@startlink[1]{}%
\providecommand \@@endlink[0]{}%
\providecommand \url  [0]{\begingroup\@sanitize@url \@url }%
\providecommand \@url [1]{\endgroup\@href {#1}{\urlprefix }}%
\providecommand \urlprefix  [0]{URL }%
\providecommand \Eprint [0]{\href }%
\providecommand \doibase [0]{http://dx.doi.org/}%
\providecommand \selectlanguage [0]{\@gobble}%
\providecommand \bibinfo  [0]{\@secondoftwo}%
\providecommand \bibfield  [0]{\@secondoftwo}%
\providecommand \translation [1]{[#1]}%
\providecommand \BibitemOpen [0]{}%
\providecommand \bibitemStop [0]{}%
\providecommand \bibitemNoStop [0]{.\EOS\space}%
\providecommand \EOS [0]{\spacefactor3000\relax}%
\providecommand \BibitemShut  [1]{\csname bibitem#1\endcsname}%
\let\auto@bib@innerbib\@empty
\bibitem [{\citenamefont {Abbott}\ \emph {et~al.}(2016)\citenamefont {Abbott}
  \emph {et~al.}}]{Abbott:2016blz}%
  \BibitemOpen
  \bibfield  {author} {\bibinfo {author} {\bibfnamefont {B.}~\bibnamefont
  {Abbott}} \emph {et~al.} (\bibinfo {collaboration} {LIGO Scientific,
  Virgo}),\ }\href {\doibase 10.1103/PhysRevLett.116.061102} {\bibfield
  {journal} {\bibinfo  {journal} {Phys. Rev. Lett.}\ }\textbf {\bibinfo
  {volume} {116}},\ \bibinfo {pages} {061102} (\bibinfo {year} {2016})},\
  \Eprint {http://arxiv.org/abs/1602.03837} {arXiv:1602.03837 [gr-qc]}
  \BibitemShut {NoStop}%
\bibitem [{\citenamefont {Janssen}\ \emph {et~al.}(2015)\citenamefont {Janssen}
  \emph {et~al.}}]{Janssen:2014dka}%
  \BibitemOpen
  \bibfield  {author} {\bibinfo {author} {\bibfnamefont {G.}~\bibnamefont
  {Janssen}} \emph {et~al.},\ }\href {\doibase 10.22323/1.215.0037} {\bibfield
  {journal} {\bibinfo  {journal} {PoS}\ }\textbf {\bibinfo {volume}
  {AASKA14}},\ \bibinfo {pages} {037} (\bibinfo {year} {2015})},\ \Eprint
  {http://arxiv.org/abs/1501.00127} {arXiv:1501.00127 [astro-ph.IM]}
  \BibitemShut {NoStop}%
\bibitem [{\citenamefont {Audley}\ \emph {et~al.}(2017)\citenamefont {Audley}
  \emph {et~al.}}]{Audley:2017drz}%
  \BibitemOpen
  \bibfield  {author} {\bibinfo {author} {\bibfnamefont {H.}~\bibnamefont
  {Audley}} \emph {et~al.} (\bibinfo {collaboration} {LISA}),\ }\href@noop {}
  {\  (\bibinfo {year} {2017})},\ \Eprint {http://arxiv.org/abs/1702.00786}
  {arXiv:1702.00786 [astro-ph.IM]} \BibitemShut {NoStop}%
\bibitem [{\citenamefont {Graham}\ \emph {et~al.}(2016)\citenamefont {Graham},
  \citenamefont {Hogan}, \citenamefont {Kasevich},\ and\ \citenamefont
  {Rajendran}}]{Graham:2016plp}%
  \BibitemOpen
  \bibfield  {author} {\bibinfo {author} {\bibfnamefont {P.~W.}\ \bibnamefont
  {Graham}}, \bibinfo {author} {\bibfnamefont {J.~M.}\ \bibnamefont {Hogan}},
  \bibinfo {author} {\bibfnamefont {M.~A.}\ \bibnamefont {Kasevich}}, \ and\
  \bibinfo {author} {\bibfnamefont {S.}~\bibnamefont {Rajendran}},\ }\href
  {\doibase 10.1103/PhysRevD.94.104022} {\bibfield  {journal} {\bibinfo
  {journal} {Phys. Rev.}\ }\textbf {\bibinfo {volume} {D94}},\ \bibinfo {pages}
  {104022} (\bibinfo {year} {2016})},\ \Eprint
  {http://arxiv.org/abs/1606.01860} {arXiv:1606.01860 [physics.atom-ph]}
  \BibitemShut {NoStop}%
\bibitem [{\citenamefont {Graham}\ \emph {et~al.}(2017)\citenamefont {Graham},
  \citenamefont {Hogan}, \citenamefont {Kasevich}, \citenamefont {Rajendran},\
  and\ \citenamefont {Romani}}]{Graham:2017pmn}%
  \BibitemOpen
  \bibfield  {author} {\bibinfo {author} {\bibfnamefont {P.~W.}\ \bibnamefont
  {Graham}}, \bibinfo {author} {\bibfnamefont {J.~M.}\ \bibnamefont {Hogan}},
  \bibinfo {author} {\bibfnamefont {M.~A.}\ \bibnamefont {Kasevich}}, \bibinfo
  {author} {\bibfnamefont {S.}~\bibnamefont {Rajendran}}, \ and\ \bibinfo
  {author} {\bibfnamefont {R.~W.}\ \bibnamefont {Romani}} (\bibinfo
  {collaboration} {MAGIS}),\ }\href@noop {} {\  (\bibinfo {year} {2017})},\
  \Eprint {http://arxiv.org/abs/1711.02225} {arXiv:1711.02225 [astro-ph.IM]}
  \BibitemShut {NoStop}%
\bibitem [{\citenamefont {Badurina}\ \emph {et~al.}(2020)\citenamefont
  {Badurina} \emph {et~al.}}]{Badurina:2019hst}%
  \BibitemOpen
  \bibfield  {author} {\bibinfo {author} {\bibfnamefont {L.}~\bibnamefont
  {Badurina}} \emph {et~al.},\ }\href {\doibase 10.1088/1475-7516/2020/05/011}
  {\bibfield  {journal} {\bibinfo  {journal} {JCAP}\ }\textbf {\bibinfo
  {volume} {05}},\ \bibinfo {pages} {011} (\bibinfo {year} {2020})},\ \Eprint
  {http://arxiv.org/abs/1911.11755} {arXiv:1911.11755 [astro-ph.CO]}
  \BibitemShut {NoStop}%
\bibitem [{\citenamefont {El-Neaj}\ \emph {et~al.}(2020)\citenamefont {El-Neaj}
  \emph {et~al.}}]{Bertoldi:2019tck}%
  \BibitemOpen
  \bibfield  {author} {\bibinfo {author} {\bibfnamefont {Y.~A.}\ \bibnamefont
  {El-Neaj}} \emph {et~al.} (\bibinfo {collaboration} {AEDGE}),\ }\href
  {\doibase 10.1140/epjqt/s40507-020-0080-0} {\bibfield  {journal} {\bibinfo
  {journal} {EPJ Quant. Technol.}\ }\textbf {\bibinfo {volume} {7}},\ \bibinfo
  {pages} {6} (\bibinfo {year} {2020})},\ \Eprint
  {http://arxiv.org/abs/1908.00802} {arXiv:1908.00802 [gr-qc]} \BibitemShut
  {NoStop}%
\bibitem [{\citenamefont {Punturo}\ \emph {et~al.}(2010)\citenamefont {Punturo}
  \emph {et~al.}}]{Punturo:2010zz}%
  \BibitemOpen
  \bibfield  {author} {\bibinfo {author} {\bibfnamefont {M.}~\bibnamefont
  {Punturo}} \emph {et~al.},\ }\href {\doibase 10.1088/0264-9381/27/19/194002}
  {\bibfield  {journal} {\bibinfo  {journal} {Class. Quant. Grav.}\ }\textbf
  {\bibinfo {volume} {27}},\ \bibinfo {pages} {194002} (\bibinfo {year}
  {2010})}\BibitemShut {NoStop}%
\bibitem [{\citenamefont {Hild}\ \emph {et~al.}(2011)\citenamefont {Hild} \emph
  {et~al.}}]{Hild:2010id}%
  \BibitemOpen
  \bibfield  {author} {\bibinfo {author} {\bibfnamefont {S.}~\bibnamefont
  {Hild}} \emph {et~al.},\ }\href {\doibase 10.1088/0264-9381/28/9/094013}
  {\bibfield  {journal} {\bibinfo  {journal} {Class. Quant. Grav.}\ }\textbf
  {\bibinfo {volume} {28}},\ \bibinfo {pages} {094013} (\bibinfo {year}
  {2011})},\ \Eprint {http://arxiv.org/abs/1012.0908} {arXiv:1012.0908 [gr-qc]}
  \BibitemShut {NoStop}%
\bibitem [{\citenamefont {Witten}(1984)}]{Witten:1984rs}%
  \BibitemOpen
  \bibfield  {author} {\bibinfo {author} {\bibfnamefont {E.}~\bibnamefont
  {Witten}},\ }\href {\doibase 10.1103/PhysRevD.30.272} {\bibfield  {journal}
  {\bibinfo  {journal} {Phys. Rev.}\ }\textbf {\bibinfo {volume} {D30}},\
  \bibinfo {pages} {272} (\bibinfo {year} {1984})}\BibitemShut {NoStop}%
\bibitem [{\citenamefont {Caprini}\ \emph {et~al.}(2016)\citenamefont {Caprini}
  \emph {et~al.}}]{Caprini:2015zlo}%
  \BibitemOpen
  \bibfield  {author} {\bibinfo {author} {\bibfnamefont {C.}~\bibnamefont
  {Caprini}} \emph {et~al.},\ }\href {\doibase 10.1088/1475-7516/2016/04/001}
  {\bibfield  {journal} {\bibinfo  {journal} {JCAP}\ }\textbf {\bibinfo
  {volume} {1604}},\ \bibinfo {pages} {001} (\bibinfo {year} {2016})},\ \Eprint
  {http://arxiv.org/abs/1512.06239} {arXiv:1512.06239 [astro-ph.CO]}
  \BibitemShut {NoStop}%
\bibitem [{\citenamefont {Caprini}\ \emph {et~al.}(2020)\citenamefont {Caprini}
  \emph {et~al.}}]{Caprini:2019egz}%
  \BibitemOpen
  \bibfield  {author} {\bibinfo {author} {\bibfnamefont {C.}~\bibnamefont
  {Caprini}} \emph {et~al.},\ }\href {\doibase 10.1088/1475-7516/2020/03/024}
  {\bibfield  {journal} {\bibinfo  {journal} {JCAP}\ }\textbf {\bibinfo
  {volume} {2003}},\ \bibinfo {pages} {024} (\bibinfo {year} {2020})},\ \Eprint
  {http://arxiv.org/abs/1910.13125} {arXiv:1910.13125 [astro-ph.CO]}
  \BibitemShut {NoStop}%
\bibitem [{\citenamefont {Coleman}(1977)}]{Coleman:1977py}%
  \BibitemOpen
  \bibfield  {author} {\bibinfo {author} {\bibfnamefont {S.~R.}\ \bibnamefont
  {Coleman}},\ }\href {\doibase 10.1103/PhysRevD.15.2929,
  10.1103/PhysRevD.16.1248} {\bibfield  {journal} {\bibinfo  {journal} {Phys.
  Rev.}\ }\textbf {\bibinfo {volume} {D15}},\ \bibinfo {pages} {2929} (\bibinfo
  {year} {1977})},\ \bibinfo {note} {[Erratum: Phys.
  Rev.D16,1248(1977)]}\BibitemShut {NoStop}%
\bibitem [{\citenamefont {Callan}\ and\ \citenamefont
  {Coleman}(1977)}]{Callan:1977pt}%
  \BibitemOpen
  \bibfield  {author} {\bibinfo {author} {\bibfnamefont {C.~G.}\ \bibnamefont
  {Callan}, \bibfnamefont {Jr.}}\ and\ \bibinfo {author} {\bibfnamefont
  {S.~R.}\ \bibnamefont {Coleman}},\ }\href {\doibase 10.1103/PhysRevD.16.1762}
  {\bibfield  {journal} {\bibinfo  {journal} {Phys. Rev.}\ }\textbf {\bibinfo
  {volume} {D16}},\ \bibinfo {pages} {1762} (\bibinfo {year}
  {1977})}\BibitemShut {NoStop}%
\bibitem [{\citenamefont {Linde}(1983)}]{Linde:1981zj}%
  \BibitemOpen
  \bibfield  {author} {\bibinfo {author} {\bibfnamefont {A.~D.}\ \bibnamefont
  {Linde}},\ }\href {\doibase 10.1016/0550-3213(83)90293-6,
  10.1016/0550-3213(83)90072-X} {\bibfield  {journal} {\bibinfo  {journal}
  {Nucl. Phys.}\ }\textbf {\bibinfo {volume} {B216}},\ \bibinfo {pages} {421}
  (\bibinfo {year} {1983})},\ \bibinfo {note} {[Erratum: Nucl.
  Phys.B223,544(1983)]}\BibitemShut {NoStop}%
\bibitem [{\citenamefont {Kosowsky}\ and\ \citenamefont
  {Turner}(1993)}]{Kosowsky:1992vn}%
  \BibitemOpen
  \bibfield  {author} {\bibinfo {author} {\bibfnamefont {A.}~\bibnamefont
  {Kosowsky}}\ and\ \bibinfo {author} {\bibfnamefont {M.~S.}\ \bibnamefont
  {Turner}},\ }\href {\doibase 10.1103/PhysRevD.47.4372} {\bibfield  {journal}
  {\bibinfo  {journal} {Phys. Rev.}\ }\textbf {\bibinfo {volume} {D47}},\
  \bibinfo {pages} {4372} (\bibinfo {year} {1993})},\ \Eprint
  {http://arxiv.org/abs/astro-ph/9211004} {arXiv:astro-ph/9211004 [astro-ph]}
  \BibitemShut {NoStop}%
\bibitem [{\citenamefont {Kamionkowski}\ \emph {et~al.}(1994)\citenamefont
  {Kamionkowski}, \citenamefont {Kosowsky},\ and\ \citenamefont
  {Turner}}]{Kamionkowski:1993fg}%
  \BibitemOpen
  \bibfield  {author} {\bibinfo {author} {\bibfnamefont {M.}~\bibnamefont
  {Kamionkowski}}, \bibinfo {author} {\bibfnamefont {A.}~\bibnamefont
  {Kosowsky}}, \ and\ \bibinfo {author} {\bibfnamefont {M.~S.}\ \bibnamefont
  {Turner}},\ }\href {\doibase 10.1103/PhysRevD.49.2837} {\bibfield  {journal}
  {\bibinfo  {journal} {Phys. Rev. D}\ }\textbf {\bibinfo {volume} {49}},\
  \bibinfo {pages} {2837} (\bibinfo {year} {1994})},\ \Eprint
  {http://arxiv.org/abs/astro-ph/9310044} {arXiv:astro-ph/9310044} \BibitemShut
  {NoStop}%
\bibitem [{\citenamefont {Bodeker}\ and\ \citenamefont
  {Moore}(2009)}]{Bodeker:2009qy}%
  \BibitemOpen
  \bibfield  {author} {\bibinfo {author} {\bibfnamefont {D.}~\bibnamefont
  {Bodeker}}\ and\ \bibinfo {author} {\bibfnamefont {G.~D.}\ \bibnamefont
  {Moore}},\ }\href {\doibase 10.1088/1475-7516/2009/05/009} {\bibfield
  {journal} {\bibinfo  {journal} {JCAP}\ }\textbf {\bibinfo {volume} {05}},\
  \bibinfo {pages} {009} (\bibinfo {year} {2009})},\ \Eprint
  {http://arxiv.org/abs/0903.4099} {arXiv:0903.4099 [hep-ph]} \BibitemShut
  {NoStop}%
\bibitem [{\citenamefont {Bodeker}\ and\ \citenamefont
  {Moore}(2017)}]{Bodeker:2017cim}%
  \BibitemOpen
  \bibfield  {author} {\bibinfo {author} {\bibfnamefont {D.}~\bibnamefont
  {Bodeker}}\ and\ \bibinfo {author} {\bibfnamefont {G.~D.}\ \bibnamefont
  {Moore}},\ }\href {\doibase 10.1088/1475-7516/2017/05/025} {\bibfield
  {journal} {\bibinfo  {journal} {JCAP}\ }\textbf {\bibinfo {volume} {05}},\
  \bibinfo {pages} {025} (\bibinfo {year} {2017})},\ \Eprint
  {http://arxiv.org/abs/1703.08215} {arXiv:1703.08215 [hep-ph]} \BibitemShut
  {NoStop}%
\bibitem [{\citenamefont {Ellis}\ \emph
  {et~al.}(2019{\natexlab{a}})\citenamefont {Ellis}, \citenamefont {Lewicki},
  \citenamefont {No},\ and\ \citenamefont {Vaskonen}}]{Ellis:2019oqb}%
  \BibitemOpen
  \bibfield  {author} {\bibinfo {author} {\bibfnamefont {J.}~\bibnamefont
  {Ellis}}, \bibinfo {author} {\bibfnamefont {M.}~\bibnamefont {Lewicki}},
  \bibinfo {author} {\bibfnamefont {J.~M.}\ \bibnamefont {No}}, \ and\ \bibinfo
  {author} {\bibfnamefont {V.}~\bibnamefont {Vaskonen}},\ }\href {\doibase
  10.1088/1475-7516/2019/06/024} {\bibfield  {journal} {\bibinfo  {journal}
  {JCAP}\ }\textbf {\bibinfo {volume} {1906}},\ \bibinfo {pages} {024}
  (\bibinfo {year} {2019}{\natexlab{a}})},\ \Eprint
  {http://arxiv.org/abs/1903.09642} {arXiv:1903.09642 [hep-ph]} \BibitemShut
  {NoStop}%
\bibitem [{\citenamefont {Huber}\ and\ \citenamefont
  {Konstandin}(2008)}]{Huber:2008hg}%
  \BibitemOpen
  \bibfield  {author} {\bibinfo {author} {\bibfnamefont {S.~J.}\ \bibnamefont
  {Huber}}\ and\ \bibinfo {author} {\bibfnamefont {T.}~\bibnamefont
  {Konstandin}},\ }\href {\doibase 10.1088/1475-7516/2008/09/022} {\bibfield
  {journal} {\bibinfo  {journal} {JCAP}\ }\textbf {\bibinfo {volume} {09}},\
  \bibinfo {pages} {022} (\bibinfo {year} {2008})},\ \Eprint
  {http://arxiv.org/abs/0806.1828} {arXiv:0806.1828 [hep-ph]} \BibitemShut
  {NoStop}%
\bibitem [{\citenamefont {Weir}(2016)}]{Weir:2016tov}%
  \BibitemOpen
  \bibfield  {author} {\bibinfo {author} {\bibfnamefont {D.~J.}\ \bibnamefont
  {Weir}},\ }\href {\doibase 10.1103/PhysRevD.93.124037} {\bibfield  {journal}
  {\bibinfo  {journal} {Phys. Rev. D}\ }\textbf {\bibinfo {volume} {93}},\
  \bibinfo {pages} {124037} (\bibinfo {year} {2016})},\ \Eprint
  {http://arxiv.org/abs/1604.08429} {arXiv:1604.08429 [astro-ph.CO]}
  \BibitemShut {NoStop}%
\bibitem [{\citenamefont {Konstandin}(2018)}]{Konstandin:2017sat}%
  \BibitemOpen
  \bibfield  {author} {\bibinfo {author} {\bibfnamefont {T.}~\bibnamefont
  {Konstandin}},\ }\href {\doibase 10.1088/1475-7516/2018/03/047} {\bibfield
  {journal} {\bibinfo  {journal} {JCAP}\ }\textbf {\bibinfo {volume} {03}},\
  \bibinfo {pages} {047} (\bibinfo {year} {2018})},\ \Eprint
  {http://arxiv.org/abs/1712.06869} {arXiv:1712.06869 [astro-ph.CO]}
  \BibitemShut {NoStop}%
\bibitem [{\citenamefont {Jinno}\ and\ \citenamefont
  {Takimoto}(2019)}]{Jinno:2017fby}%
  \BibitemOpen
  \bibfield  {author} {\bibinfo {author} {\bibfnamefont {R.}~\bibnamefont
  {Jinno}}\ and\ \bibinfo {author} {\bibfnamefont {M.}~\bibnamefont
  {Takimoto}},\ }\href {\doibase 10.1088/1475-7516/2019/01/060} {\bibfield
  {journal} {\bibinfo  {journal} {JCAP}\ }\textbf {\bibinfo {volume} {01}},\
  \bibinfo {pages} {060} (\bibinfo {year} {2019})},\ \Eprint
  {http://arxiv.org/abs/1707.03111} {arXiv:1707.03111 [hep-ph]} \BibitemShut
  {NoStop}%
\bibitem [{\citenamefont {Child}\ and\ \citenamefont
  {Giblin}(2012)}]{Child:2012qg}%
  \BibitemOpen
  \bibfield  {author} {\bibinfo {author} {\bibfnamefont {H.~L.}\ \bibnamefont
  {Child}}\ and\ \bibinfo {author} {\bibfnamefont {J.}~\bibnamefont {Giblin},
  \bibfnamefont {John~T.}},\ }\href {\doibase 10.1088/1475-7516/2012/10/001}
  {\bibfield  {journal} {\bibinfo  {journal} {JCAP}\ }\textbf {\bibinfo
  {volume} {10}},\ \bibinfo {pages} {001} (\bibinfo {year} {2012})},\ \Eprint
  {http://arxiv.org/abs/1207.6408} {arXiv:1207.6408 [astro-ph.CO]} \BibitemShut
  {NoStop}%
\bibitem [{\citenamefont {Cutting}\ \emph {et~al.}(2018)\citenamefont
  {Cutting}, \citenamefont {Hindmarsh},\ and\ \citenamefont
  {Weir}}]{Cutting:2018tjt}%
  \BibitemOpen
  \bibfield  {author} {\bibinfo {author} {\bibfnamefont {D.}~\bibnamefont
  {Cutting}}, \bibinfo {author} {\bibfnamefont {M.}~\bibnamefont {Hindmarsh}},
  \ and\ \bibinfo {author} {\bibfnamefont {D.~J.}\ \bibnamefont {Weir}},\
  }\href {\doibase 10.1103/PhysRevD.97.123513} {\bibfield  {journal} {\bibinfo
  {journal} {Phys. Rev.}\ }\textbf {\bibinfo {volume} {D97}},\ \bibinfo {pages}
  {123513} (\bibinfo {year} {2018})},\ \Eprint
  {http://arxiv.org/abs/1802.05712} {arXiv:1802.05712 [astro-ph.CO]}
  \BibitemShut {NoStop}%
\bibitem [{\citenamefont {Cutting}\ \emph {et~al.}(2020)\citenamefont
  {Cutting}, \citenamefont {Escartin}, \citenamefont {Hindmarsh},\ and\
  \citenamefont {Weir}}]{Cutting:2020nla}%
  \BibitemOpen
  \bibfield  {author} {\bibinfo {author} {\bibfnamefont {D.}~\bibnamefont
  {Cutting}}, \bibinfo {author} {\bibfnamefont {E.~G.}\ \bibnamefont
  {Escartin}}, \bibinfo {author} {\bibfnamefont {M.}~\bibnamefont {Hindmarsh}},
  \ and\ \bibinfo {author} {\bibfnamefont {D.~J.}\ \bibnamefont {Weir}},\
  }\href@noop {} {\  (\bibinfo {year} {2020})},\ \Eprint
  {http://arxiv.org/abs/2005.13537} {arXiv:2005.13537 [astro-ph.CO]}
  \BibitemShut {NoStop}%
\bibitem [{\citenamefont {Huber}\ \emph {et~al.}(2016)\citenamefont {Huber},
  \citenamefont {Konstandin}, \citenamefont {Nardini},\ and\ \citenamefont
  {Rues}}]{Huber:2015znp}%
  \BibitemOpen
  \bibfield  {author} {\bibinfo {author} {\bibfnamefont {S.~J.}\ \bibnamefont
  {Huber}}, \bibinfo {author} {\bibfnamefont {T.}~\bibnamefont {Konstandin}},
  \bibinfo {author} {\bibfnamefont {G.}~\bibnamefont {Nardini}}, \ and\
  \bibinfo {author} {\bibfnamefont {I.}~\bibnamefont {Rues}},\ }\href {\doibase
  10.1088/1475-7516/2016/03/036} {\bibfield  {journal} {\bibinfo  {journal}
  {JCAP}\ }\textbf {\bibinfo {volume} {03}},\ \bibinfo {pages} {036} (\bibinfo
  {year} {2016})},\ \Eprint {http://arxiv.org/abs/1512.06357} {arXiv:1512.06357
  [hep-ph]} \BibitemShut {NoStop}%
\bibitem [{\citenamefont {Jinno}\ and\ \citenamefont
  {Takimoto}(2017)}]{Jinno:2016knw}%
  \BibitemOpen
  \bibfield  {author} {\bibinfo {author} {\bibfnamefont {R.}~\bibnamefont
  {Jinno}}\ and\ \bibinfo {author} {\bibfnamefont {M.}~\bibnamefont
  {Takimoto}},\ }\href {\doibase 10.1103/PhysRevD.95.015020} {\bibfield
  {journal} {\bibinfo  {journal} {Phys. Rev. D}\ }\textbf {\bibinfo {volume}
  {95}},\ \bibinfo {pages} {015020} (\bibinfo {year} {2017})},\ \Eprint
  {http://arxiv.org/abs/1604.05035} {arXiv:1604.05035 [hep-ph]} \BibitemShut
  {NoStop}%
\bibitem [{\citenamefont {Iso}\ \emph {et~al.}(2017)\citenamefont {Iso},
  \citenamefont {Serpico},\ and\ \citenamefont {Shimada}}]{Iso:2017uuu}%
  \BibitemOpen
  \bibfield  {author} {\bibinfo {author} {\bibfnamefont {S.}~\bibnamefont
  {Iso}}, \bibinfo {author} {\bibfnamefont {P.~D.}\ \bibnamefont {Serpico}}, \
  and\ \bibinfo {author} {\bibfnamefont {K.}~\bibnamefont {Shimada}},\ }\href
  {\doibase 10.1103/PhysRevLett.119.141301} {\bibfield  {journal} {\bibinfo
  {journal} {Phys. Rev. Lett.}\ }\textbf {\bibinfo {volume} {119}},\ \bibinfo
  {pages} {141301} (\bibinfo {year} {2017})},\ \Eprint
  {http://arxiv.org/abs/1704.04955} {arXiv:1704.04955 [hep-ph]} \BibitemShut
  {NoStop}%
\bibitem [{\citenamefont {Demidov}\ \emph {et~al.}(2018)\citenamefont
  {Demidov}, \citenamefont {Gorbunov},\ and\ \citenamefont
  {Kirpichnikov}}]{Demidov:2017lzf}%
  \BibitemOpen
  \bibfield  {author} {\bibinfo {author} {\bibfnamefont {S.}~\bibnamefont
  {Demidov}}, \bibinfo {author} {\bibfnamefont {D.}~\bibnamefont {Gorbunov}}, \
  and\ \bibinfo {author} {\bibfnamefont {D.}~\bibnamefont {Kirpichnikov}},\
  }\href {\doibase 10.1016/j.physletb.2018.02.007} {\bibfield  {journal}
  {\bibinfo  {journal} {Phys. Lett. B}\ }\textbf {\bibinfo {volume} {779}},\
  \bibinfo {pages} {191} (\bibinfo {year} {2018})},\ \Eprint
  {http://arxiv.org/abs/1712.00087} {arXiv:1712.00087 [hep-ph]} \BibitemShut
  {NoStop}%
\bibitem [{\citenamefont {Hashino}\ \emph {et~al.}(2018)\citenamefont
  {Hashino}, \citenamefont {Kakizaki}, \citenamefont {Kanemura}, \citenamefont
  {Ko},\ and\ \citenamefont {Matsui}}]{Hashino:2018zsi}%
  \BibitemOpen
  \bibfield  {author} {\bibinfo {author} {\bibfnamefont {K.}~\bibnamefont
  {Hashino}}, \bibinfo {author} {\bibfnamefont {M.}~\bibnamefont {Kakizaki}},
  \bibinfo {author} {\bibfnamefont {S.}~\bibnamefont {Kanemura}}, \bibinfo
  {author} {\bibfnamefont {P.}~\bibnamefont {Ko}}, \ and\ \bibinfo {author}
  {\bibfnamefont {T.}~\bibnamefont {Matsui}},\ }\href {\doibase
  10.1007/JHEP06(2018)088} {\bibfield  {journal} {\bibinfo  {journal} {JHEP}\
  }\textbf {\bibinfo {volume} {06}},\ \bibinfo {pages} {088} (\bibinfo {year}
  {2018})},\ \Eprint {http://arxiv.org/abs/1802.02947} {arXiv:1802.02947
  [hep-ph]} \BibitemShut {NoStop}%
\bibitem [{\citenamefont {Marzo}\ \emph {et~al.}(2019)\citenamefont {Marzo},
  \citenamefont {Marzola},\ and\ \citenamefont {Vaskonen}}]{Marzo:2018nov}%
  \BibitemOpen
  \bibfield  {author} {\bibinfo {author} {\bibfnamefont {C.}~\bibnamefont
  {Marzo}}, \bibinfo {author} {\bibfnamefont {L.}~\bibnamefont {Marzola}}, \
  and\ \bibinfo {author} {\bibfnamefont {V.}~\bibnamefont {Vaskonen}},\ }\href
  {\doibase 10.1140/epjc/s10052-019-7076-x} {\bibfield  {journal} {\bibinfo
  {journal} {Eur. Phys. J.}\ }\textbf {\bibinfo {volume} {C79}},\ \bibinfo
  {pages} {601} (\bibinfo {year} {2019})},\ \Eprint
  {http://arxiv.org/abs/1811.11169} {arXiv:1811.11169 [hep-ph]} \BibitemShut
  {NoStop}%
\bibitem [{\citenamefont {Miura}\ \emph {et~al.}(2019)\citenamefont {Miura},
  \citenamefont {Ohki}, \citenamefont {Otani},\ and\ \citenamefont
  {Yamawaki}}]{Miura:2018dsy}%
  \BibitemOpen
  \bibfield  {author} {\bibinfo {author} {\bibfnamefont {K.}~\bibnamefont
  {Miura}}, \bibinfo {author} {\bibfnamefont {H.}~\bibnamefont {Ohki}},
  \bibinfo {author} {\bibfnamefont {S.}~\bibnamefont {Otani}}, \ and\ \bibinfo
  {author} {\bibfnamefont {K.}~\bibnamefont {Yamawaki}},\ }\href {\doibase
  10.1007/JHEP10(2019)194} {\bibfield  {journal} {\bibinfo  {journal} {JHEP}\
  }\textbf {\bibinfo {volume} {10}},\ \bibinfo {pages} {194} (\bibinfo {year}
  {2019})},\ \Eprint {http://arxiv.org/abs/1811.05670} {arXiv:1811.05670
  [hep-ph]} \BibitemShut {NoStop}%
\bibitem [{\citenamefont {Azatov}\ \emph {et~al.}(2019)\citenamefont {Azatov},
  \citenamefont {Barducci},\ and\ \citenamefont {Sgarlata}}]{Azatov:2019png}%
  \BibitemOpen
  \bibfield  {author} {\bibinfo {author} {\bibfnamefont {A.}~\bibnamefont
  {Azatov}}, \bibinfo {author} {\bibfnamefont {D.}~\bibnamefont {Barducci}}, \
  and\ \bibinfo {author} {\bibfnamefont {F.}~\bibnamefont {Sgarlata}},\
  }\href@noop {} {\  (\bibinfo {year} {2019})},\ \Eprint
  {http://arxiv.org/abs/1910.01124} {arXiv:1910.01124 [hep-ph]} \BibitemShut
  {NoStop}%
\bibitem [{\citenamefont {Ellis}\ \emph
  {et~al.}(2019{\natexlab{b}})\citenamefont {Ellis}, \citenamefont {Lewicki},\
  and\ \citenamefont {No}}]{Ellis:2018mja}%
  \BibitemOpen
  \bibfield  {author} {\bibinfo {author} {\bibfnamefont {J.}~\bibnamefont
  {Ellis}}, \bibinfo {author} {\bibfnamefont {M.}~\bibnamefont {Lewicki}}, \
  and\ \bibinfo {author} {\bibfnamefont {J.~M.}\ \bibnamefont {No}},\ }\href
  {\doibase 10.1088/1475-7516/2019/04/003} {\bibfield  {journal} {\bibinfo
  {journal} {JCAP}\ }\textbf {\bibinfo {volume} {04}},\ \bibinfo {pages} {003}
  (\bibinfo {year} {2019}{\natexlab{b}})},\ \Eprint
  {http://arxiv.org/abs/1809.08242} {arXiv:1809.08242 [hep-ph]} \BibitemShut
  {NoStop}%
\bibitem [{\citenamefont {Randall}\ and\ \citenamefont
  {Servant}(2007)}]{Randall:2006py}%
  \BibitemOpen
  \bibfield  {author} {\bibinfo {author} {\bibfnamefont {L.}~\bibnamefont
  {Randall}}\ and\ \bibinfo {author} {\bibfnamefont {G.}~\bibnamefont
  {Servant}},\ }\href {\doibase 10.1088/1126-6708/2007/05/054} {\bibfield
  {journal} {\bibinfo  {journal} {JHEP}\ }\textbf {\bibinfo {volume} {05}},\
  \bibinfo {pages} {054} (\bibinfo {year} {2007})},\ \Eprint
  {http://arxiv.org/abs/hep-ph/0607158} {arXiv:hep-ph/0607158 [hep-ph]}
  \BibitemShut {NoStop}%
\bibitem [{\citenamefont {Konstandin}\ and\ \citenamefont
  {Servant}(2011{\natexlab{a}})}]{Konstandin:2011dr}%
  \BibitemOpen
  \bibfield  {author} {\bibinfo {author} {\bibfnamefont {T.}~\bibnamefont
  {Konstandin}}\ and\ \bibinfo {author} {\bibfnamefont {G.}~\bibnamefont
  {Servant}},\ }\href {\doibase 10.1088/1475-7516/2011/12/009} {\bibfield
  {journal} {\bibinfo  {journal} {JCAP}\ }\textbf {\bibinfo {volume} {1112}},\
  \bibinfo {pages} {009} (\bibinfo {year} {2011}{\natexlab{a}})},\ \Eprint
  {http://arxiv.org/abs/1104.4791} {arXiv:1104.4791 [hep-ph]} \BibitemShut
  {NoStop}%
\bibitem [{\citenamefont {Konstandin}\ and\ \citenamefont
  {Servant}(2011{\natexlab{b}})}]{Konstandin:2011ds}%
  \BibitemOpen
  \bibfield  {author} {\bibinfo {author} {\bibfnamefont {T.}~\bibnamefont
  {Konstandin}}\ and\ \bibinfo {author} {\bibfnamefont {G.}~\bibnamefont
  {Servant}},\ }\href {\doibase 10.1088/1475-7516/2011/07/024} {\bibfield
  {journal} {\bibinfo  {journal} {JCAP}\ }\textbf {\bibinfo {volume} {07}},\
  \bibinfo {pages} {024} (\bibinfo {year} {2011}{\natexlab{b}})},\ \Eprint
  {http://arxiv.org/abs/1104.4793} {arXiv:1104.4793 [hep-ph]} \BibitemShut
  {NoStop}%
\bibitem [{\citenamefont {von Harling}\ and\ \citenamefont
  {Servant}(2018)}]{vonHarling:2017yew}%
  \BibitemOpen
  \bibfield  {author} {\bibinfo {author} {\bibfnamefont {B.}~\bibnamefont {von
  Harling}}\ and\ \bibinfo {author} {\bibfnamefont {G.}~\bibnamefont
  {Servant}},\ }\href {\doibase 10.1007/JHEP01(2018)159} {\bibfield  {journal}
  {\bibinfo  {journal} {JHEP}\ }\textbf {\bibinfo {volume} {01}},\ \bibinfo
  {pages} {159} (\bibinfo {year} {2018})},\ \Eprint
  {http://arxiv.org/abs/1711.11554} {arXiv:1711.11554 [hep-ph]} \BibitemShut
  {NoStop}%
\bibitem [{\citenamefont {Kobakhidze}\ \emph {et~al.}(2017)\citenamefont
  {Kobakhidze}, \citenamefont {Lagger}, \citenamefont {Manning},\ and\
  \citenamefont {Yue}}]{Kobakhidze:2017mru}%
  \BibitemOpen
  \bibfield  {author} {\bibinfo {author} {\bibfnamefont {A.}~\bibnamefont
  {Kobakhidze}}, \bibinfo {author} {\bibfnamefont {C.}~\bibnamefont {Lagger}},
  \bibinfo {author} {\bibfnamefont {A.}~\bibnamefont {Manning}}, \ and\
  \bibinfo {author} {\bibfnamefont {J.}~\bibnamefont {Yue}},\ }\href {\doibase
  10.1140/epjc/s10052-017-5132-y} {\bibfield  {journal} {\bibinfo  {journal}
  {Eur. Phys. J. C}\ }\textbf {\bibinfo {volume} {77}},\ \bibinfo {pages} {570}
  (\bibinfo {year} {2017})},\ \Eprint {http://arxiv.org/abs/1703.06552}
  {arXiv:1703.06552 [hep-ph]} \BibitemShut {NoStop}%
\bibitem [{\citenamefont {Marzola}\ \emph {et~al.}(2017)\citenamefont
  {Marzola}, \citenamefont {Racioppi},\ and\ \citenamefont
  {Vaskonen}}]{Marzola:2017jzl}%
  \BibitemOpen
  \bibfield  {author} {\bibinfo {author} {\bibfnamefont {L.}~\bibnamefont
  {Marzola}}, \bibinfo {author} {\bibfnamefont {A.}~\bibnamefont {Racioppi}}, \
  and\ \bibinfo {author} {\bibfnamefont {V.}~\bibnamefont {Vaskonen}},\ }\href
  {\doibase 10.1140/epjc/s10052-017-4996-1} {\bibfield  {journal} {\bibinfo
  {journal} {Eur. Phys. J.}\ }\textbf {\bibinfo {volume} {C77}},\ \bibinfo
  {pages} {484} (\bibinfo {year} {2017})},\ \Eprint
  {http://arxiv.org/abs/1704.01034} {arXiv:1704.01034 [hep-ph]} \BibitemShut
  {NoStop}%
\bibitem [{\citenamefont {Prokopec}\ \emph {et~al.}(2019)\citenamefont
  {Prokopec}, \citenamefont {Rezacek},\ and\ \citenamefont
  {\'Swie\.zewska}}]{Prokopec:2018tnq}%
  \BibitemOpen
  \bibfield  {author} {\bibinfo {author} {\bibfnamefont {T.}~\bibnamefont
  {Prokopec}}, \bibinfo {author} {\bibfnamefont {J.}~\bibnamefont {Rezacek}}, \
  and\ \bibinfo {author} {\bibfnamefont {B.~a.}\ \bibnamefont
  {\'Swie\.zewska}},\ }\href {\doibase 10.1088/1475-7516/2019/02/009}
  {\bibfield  {journal} {\bibinfo  {journal} {JCAP}\ }\textbf {\bibinfo
  {volume} {02}},\ \bibinfo {pages} {009} (\bibinfo {year} {2019})},\ \Eprint
  {http://arxiv.org/abs/1809.11129} {arXiv:1809.11129 [hep-ph]} \BibitemShut
  {NoStop}%
\bibitem [{\citenamefont {Hambye}\ \emph {et~al.}(2018)\citenamefont {Hambye},
  \citenamefont {Strumia},\ and\ \citenamefont {Teresi}}]{Hambye:2018qjv}%
  \BibitemOpen
  \bibfield  {author} {\bibinfo {author} {\bibfnamefont {T.}~\bibnamefont
  {Hambye}}, \bibinfo {author} {\bibfnamefont {A.}~\bibnamefont {Strumia}}, \
  and\ \bibinfo {author} {\bibfnamefont {D.}~\bibnamefont {Teresi}},\ }\href
  {\doibase 10.1007/JHEP08(2018)188} {\bibfield  {journal} {\bibinfo  {journal}
  {JHEP}\ }\textbf {\bibinfo {volume} {08}},\ \bibinfo {pages} {188} (\bibinfo
  {year} {2018})},\ \Eprint {http://arxiv.org/abs/1805.01473} {arXiv:1805.01473
  [hep-ph]} \BibitemShut {NoStop}%
\bibitem [{\citenamefont {Baratella}\ \emph {et~al.}(2019)\citenamefont
  {Baratella}, \citenamefont {Pomarol},\ and\ \citenamefont
  {Rompineve}}]{Baratella:2018pxi}%
  \BibitemOpen
  \bibfield  {author} {\bibinfo {author} {\bibfnamefont {P.}~\bibnamefont
  {Baratella}}, \bibinfo {author} {\bibfnamefont {A.}~\bibnamefont {Pomarol}},
  \ and\ \bibinfo {author} {\bibfnamefont {F.}~\bibnamefont {Rompineve}},\
  }\href {\doibase 10.1007/JHEP03(2019)100} {\bibfield  {journal} {\bibinfo
  {journal} {JHEP}\ }\textbf {\bibinfo {volume} {03}},\ \bibinfo {pages} {100}
  (\bibinfo {year} {2019})},\ \Eprint {http://arxiv.org/abs/1812.06996}
  {arXiv:1812.06996 [hep-ph]} \BibitemShut {NoStop}%
\bibitem [{\citenamefont {Bruggisser}\ \emph {et~al.}(2018)\citenamefont
  {Bruggisser}, \citenamefont {Von~Harling}, \citenamefont {Matsedonskyi},\
  and\ \citenamefont {Servant}}]{Bruggisser:2018mrt}%
  \BibitemOpen
  \bibfield  {author} {\bibinfo {author} {\bibfnamefont {S.}~\bibnamefont
  {Bruggisser}}, \bibinfo {author} {\bibfnamefont {B.}~\bibnamefont
  {Von~Harling}}, \bibinfo {author} {\bibfnamefont {O.}~\bibnamefont
  {Matsedonskyi}}, \ and\ \bibinfo {author} {\bibfnamefont {G.}~\bibnamefont
  {Servant}},\ }\href {\doibase 10.1007/JHEP12(2018)099} {\bibfield  {journal}
  {\bibinfo  {journal} {JHEP}\ }\textbf {\bibinfo {volume} {12}},\ \bibinfo
  {pages} {099} (\bibinfo {year} {2018})},\ \Eprint
  {http://arxiv.org/abs/1804.07314} {arXiv:1804.07314 [hep-ph]} \BibitemShut
  {NoStop}%
\bibitem [{\citenamefont {Aoki}\ and\ \citenamefont
  {Kubo}(2020)}]{Aoki:2019mlt}%
  \BibitemOpen
  \bibfield  {author} {\bibinfo {author} {\bibfnamefont {M.}~\bibnamefont
  {Aoki}}\ and\ \bibinfo {author} {\bibfnamefont {J.}~\bibnamefont {Kubo}},\
  }\href {\doibase 10.1088/1475-7516/2020/04/001} {\bibfield  {journal}
  {\bibinfo  {journal} {JCAP}\ }\textbf {\bibinfo {volume} {04}},\ \bibinfo
  {pages} {001} (\bibinfo {year} {2020})},\ \Eprint
  {http://arxiv.org/abs/1910.05025} {arXiv:1910.05025 [hep-ph]} \BibitemShut
  {NoStop}%
\bibitem [{\citenamefont {Delle~Rose}\ \emph {et~al.}(2020)\citenamefont
  {Delle~Rose}, \citenamefont {Panico}, \citenamefont {Redi},\ and\
  \citenamefont {Tesi}}]{DelleRose:2019pgi}%
  \BibitemOpen
  \bibfield  {author} {\bibinfo {author} {\bibfnamefont {L.}~\bibnamefont
  {Delle~Rose}}, \bibinfo {author} {\bibfnamefont {G.}~\bibnamefont {Panico}},
  \bibinfo {author} {\bibfnamefont {M.}~\bibnamefont {Redi}}, \ and\ \bibinfo
  {author} {\bibfnamefont {A.}~\bibnamefont {Tesi}},\ }\href {\doibase
  10.1007/JHEP04(2020)025} {\bibfield  {journal} {\bibinfo  {journal} {JHEP}\
  }\textbf {\bibinfo {volume} {04}},\ \bibinfo {pages} {025} (\bibinfo {year}
  {2020})},\ \Eprint {http://arxiv.org/abs/1912.06139} {arXiv:1912.06139
  [hep-ph]} \BibitemShut {NoStop}%
\bibitem [{\citenamefont {Fujikura}\ \emph {et~al.}(2020)\citenamefont
  {Fujikura}, \citenamefont {Nakai},\ and\ \citenamefont
  {Yamada}}]{Fujikura:2019oyi}%
  \BibitemOpen
  \bibfield  {author} {\bibinfo {author} {\bibfnamefont {K.}~\bibnamefont
  {Fujikura}}, \bibinfo {author} {\bibfnamefont {Y.}~\bibnamefont {Nakai}}, \
  and\ \bibinfo {author} {\bibfnamefont {M.}~\bibnamefont {Yamada}},\ }\href
  {\doibase 10.1007/JHEP02(2020)111} {\bibfield  {journal} {\bibinfo  {journal}
  {JHEP}\ }\textbf {\bibinfo {volume} {02}},\ \bibinfo {pages} {111} (\bibinfo
  {year} {2020})},\ \Eprint {http://arxiv.org/abs/1910.07546} {arXiv:1910.07546
  [hep-ph]} \BibitemShut {NoStop}%
\bibitem [{\citenamefont {Wang}\ \emph {et~al.}(2020)\citenamefont {Wang},
  \citenamefont {Huang},\ and\ \citenamefont {Zhang}}]{Wang:2020jrd}%
  \BibitemOpen
  \bibfield  {author} {\bibinfo {author} {\bibfnamefont {X.}~\bibnamefont
  {Wang}}, \bibinfo {author} {\bibfnamefont {F.~P.}\ \bibnamefont {Huang}}, \
  and\ \bibinfo {author} {\bibfnamefont {X.}~\bibnamefont {Zhang}},\ }\href
  {\doibase 10.1088/1475-7516/2020/05/045} {\bibfield  {journal} {\bibinfo
  {journal} {JCAP}\ }\textbf {\bibinfo {volume} {05}},\ \bibinfo {pages} {045}
  (\bibinfo {year} {2020})},\ \Eprint {http://arxiv.org/abs/2003.08892}
  {arXiv:2003.08892 [hep-ph]} \BibitemShut {NoStop}%
\bibitem [{\citenamefont {Coleman}\ and\ \citenamefont
  {Weinberg}(1973)}]{Coleman:1973jx}%
  \BibitemOpen
  \bibfield  {author} {\bibinfo {author} {\bibfnamefont {S.~R.}\ \bibnamefont
  {Coleman}}\ and\ \bibinfo {author} {\bibfnamefont {E.~J.}\ \bibnamefont
  {Weinberg}},\ }\href {\doibase 10.1103/PhysRevD.7.1888} {\bibfield  {journal}
  {\bibinfo  {journal} {Phys. Rev. D}\ }\textbf {\bibinfo {volume} {7}},\
  \bibinfo {pages} {1888} (\bibinfo {year} {1973})}\BibitemShut {NoStop}%
\bibitem [{\citenamefont {Hawking}\ \emph {et~al.}(1982)\citenamefont
  {Hawking}, \citenamefont {Moss},\ and\ \citenamefont
  {Stewart}}]{Hawking:1982ga}%
  \BibitemOpen
  \bibfield  {author} {\bibinfo {author} {\bibfnamefont {S.~W.}\ \bibnamefont
  {Hawking}}, \bibinfo {author} {\bibfnamefont {I.~G.}\ \bibnamefont {Moss}}, \
  and\ \bibinfo {author} {\bibfnamefont {J.~M.}\ \bibnamefont {Stewart}},\
  }\href {\doibase 10.1103/PhysRevD.26.2681} {\bibfield  {journal} {\bibinfo
  {journal} {Phys. Rev.}\ }\textbf {\bibinfo {volume} {D26}},\ \bibinfo {pages}
  {2681} (\bibinfo {year} {1982})}\BibitemShut {NoStop}%
\bibitem [{\citenamefont {Jinno}\ \emph {et~al.}(2019)\citenamefont {Jinno},
  \citenamefont {Konstandin},\ and\ \citenamefont {Takimoto}}]{Jinno:2019bxw}%
  \BibitemOpen
  \bibfield  {author} {\bibinfo {author} {\bibfnamefont {R.}~\bibnamefont
  {Jinno}}, \bibinfo {author} {\bibfnamefont {T.}~\bibnamefont {Konstandin}}, \
  and\ \bibinfo {author} {\bibfnamefont {M.}~\bibnamefont {Takimoto}},\ }\href
  {\doibase 10.1088/1475-7516/2019/09/035} {\bibfield  {journal} {\bibinfo
  {journal} {JCAP}\ }\textbf {\bibinfo {volume} {1909}},\ \bibinfo {pages}
  {035} (\bibinfo {year} {2019})},\ \Eprint {http://arxiv.org/abs/1906.02588}
  {arXiv:1906.02588 [hep-ph]} \BibitemShut {NoStop}%
\bibitem [{\citenamefont {Lewicki}\ and\ \citenamefont
  {Vaskonen}(2020)}]{Lewicki:2019gmv}%
  \BibitemOpen
  \bibfield  {author} {\bibinfo {author} {\bibfnamefont {M.}~\bibnamefont
  {Lewicki}}\ and\ \bibinfo {author} {\bibfnamefont {V.}~\bibnamefont
  {Vaskonen}},\ }\href {\doibase https://doi.org/10.1016/j.dark.2020.100672}
  {\bibfield  {journal} {\bibinfo  {journal} {Phys. Dark Universe}\ }\textbf
  {\bibinfo {volume} {30}},\ \bibinfo {pages} {100672} (\bibinfo {year}
  {2020})},\ \Eprint {http://arxiv.org/abs/1912.00997} {arXiv:1912.00997
  [astro-ph.CO]} \BibitemShut {NoStop}%
\bibitem [{\citenamefont {Weinberg}(1972)}]{Weinberg:1972kfs}%
  \BibitemOpen
  \bibfield  {author} {\bibinfo {author} {\bibfnamefont {S.}~\bibnamefont
  {Weinberg}},\ }\href@noop {} {\emph {\bibinfo {title} {{Gravitation and
  Cosmology}}}}\ (\bibinfo  {publisher} {John Wiley and Sons},\ \bibinfo
  {address} {New York},\ \bibinfo {year} {1972})\BibitemShut {NoStop}%
\bibitem [{\citenamefont {Burden}\ and\ \citenamefont
  {Faires}(1997)}]{Burden1997}%
  \BibitemOpen
  \bibfield  {author} {\bibinfo {author} {\bibfnamefont {R.}~\bibnamefont
  {Burden}}\ and\ \bibinfo {author} {\bibfnamefont {J.}~\bibnamefont
  {Faires}},\ }\href@noop {} {\emph {\bibinfo {title} {Numerical Analysis}}},\
  Mathematics Series\ (\bibinfo  {publisher} {Brooks/Cole Publishing Company},\
  \bibinfo {year} {1997})\BibitemShut {NoStop}%
\bibitem [{\citenamefont {Aghanim}\ \emph {et~al.}(2018)\citenamefont {Aghanim}
  \emph {et~al.}}]{Aghanim:2018eyx}%
  \BibitemOpen
  \bibfield  {author} {\bibinfo {author} {\bibfnamefont {N.}~\bibnamefont
  {Aghanim}} \emph {et~al.} (\bibinfo {collaboration} {Planck}),\ }\href@noop
  {} {\  (\bibinfo {year} {2018})},\ \Eprint {http://arxiv.org/abs/1807.06209}
  {arXiv:1807.06209 [astro-ph.CO]} \BibitemShut {NoStop}%
\bibitem [{\citenamefont {Allahverdi}\ \emph {et~al.}(2020)\citenamefont
  {Allahverdi} \emph {et~al.}}]{Allahverdi:2020bys}%
  \BibitemOpen
  \bibfield  {author} {\bibinfo {author} {\bibfnamefont {R.}~\bibnamefont
  {Allahverdi}} \emph {et~al.},\ }\href@noop {} {\  (\bibinfo {year} {2020})},\
  \Eprint {http://arxiv.org/abs/2006.16182} {arXiv:2006.16182 [astro-ph.CO]}
  \BibitemShut {NoStop}%
\bibitem [{\citenamefont {Caprini}\ \emph {et~al.}(2009)\citenamefont
  {Caprini}, \citenamefont {Durrer}, \citenamefont {Konstandin},\ and\
  \citenamefont {Servant}}]{Caprini:2009fx}%
  \BibitemOpen
  \bibfield  {author} {\bibinfo {author} {\bibfnamefont {C.}~\bibnamefont
  {Caprini}}, \bibinfo {author} {\bibfnamefont {R.}~\bibnamefont {Durrer}},
  \bibinfo {author} {\bibfnamefont {T.}~\bibnamefont {Konstandin}}, \ and\
  \bibinfo {author} {\bibfnamefont {G.}~\bibnamefont {Servant}},\ }\href
  {\doibase 10.1103/PhysRevD.79.083519} {\bibfield  {journal} {\bibinfo
  {journal} {Phys. Rev. D}\ }\textbf {\bibinfo {volume} {79}},\ \bibinfo
  {pages} {083519} (\bibinfo {year} {2009})},\ \Eprint
  {http://arxiv.org/abs/0901.1661} {arXiv:0901.1661 [astro-ph.CO]} \BibitemShut
  {NoStop}%
\bibitem [{\citenamefont {Cai}\ \emph {et~al.}(2019)\citenamefont {Cai},
  \citenamefont {Pi},\ and\ \citenamefont {Sasaki}}]{Cai:2019cdl}%
  \BibitemOpen
  \bibfield  {author} {\bibinfo {author} {\bibfnamefont {R.-G.}\ \bibnamefont
  {Cai}}, \bibinfo {author} {\bibfnamefont {S.}~\bibnamefont {Pi}}, \ and\
  \bibinfo {author} {\bibfnamefont {M.}~\bibnamefont {Sasaki}},\ }\href@noop {}
  {\  (\bibinfo {year} {2019})},\ \Eprint {http://arxiv.org/abs/1909.13728}
  {arXiv:1909.13728 [astro-ph.CO]} \BibitemShut {NoStop}%
\end{thebibliography}%
\end{document}